# Elucidation of lasing mechanism in all-inorganic perovskite microcavity with exciton/free-carrier hybrid nature

Shuki Enomoto, Yusuke Ueda, Tomoya Tagami, Kentaro Fujiwara, Shun Takahashi, and Kenichi Yamashita[*]

*Faculty of Electrical Engineering and Electronics, Kyoto Institute of Technology, Matsugasaki, Sakyo-ku, Kyoto 606-8585, Japan*

*e-mail: yamasita@kit.ac.jp

Lead-halide perovskites are highly promising for various optoelectronic applications including laser devices. However, fundamental photophysics describing the lasing behaviour in this material family have not been investigated systematically, and thus the detailed lasing mechanism remains veiled. Here, we demonstrate that an all-inorganic perovskite shows a lasing emission based on a novel principle. A series of photoluminescence experiment results reveals that the excited state of $CsPbBr_3$ single crystal exhibits free carrier nature. In a form of microcavity, on the other hand, the $CsPbBr_3$ single crystal shows emission from polaritonic excited state, demonstrating the coexistence of excitonic nature. Room temperature polariton lasing is also clearly observed. This unprecedent mechanism, i.e. polariton lasing based on free-carrier reservoir, is accomplished by the exciton/free-carrier hybrid nature of $CsPbBr_3$.

## Introduction

Recent advances in research fields on lead halide perovskites unveiled great potential of this material family for applications in various optoelectronic technologies such as photoelectric conversion and electroluminescence.[1-3] Excellent optoelectronic properties of the perovskites, such as the very wide variation in bandgap[4,5] and high photoluminescence quantum yield,[6,7] offer a considerable prospect also as lasing media with various functionalities,[3,8] e.g. the colour tunability and low power consumption. Lasing operations with many types of resonators including microcavity,[6,9] distributed feedback,[10,11] and whispering gallery modes (WGM) [12-14] have been demonstrated with perovskites in forms of thin films,[6,8-11] microcrystals,[12-14] and colloidal nanocrystals.[15,16] In addition to conventional 'photon lasing', some studies have demonstrated 'polariton lasing' in the perovskite microcavities,[17,18] where an optical cavity mode and a exciton transition dipole moment are strongly coupled to form a polariton quasi-particle.[19-23] The macroscopically coherent state of condensed polariton[22-24] is an important platform to develop the devices used for quantum information science,[25] e.g. quantum computing[26,27] and single photon emission[28].

However, due to the mysterious nature of excited states of lead halide perovskites, detailed mechanism of lasing phenomena found in these materials are not yet fully understood.[29] In the first place, it is unclear whether the excited species that causes lasing is free carrier or exciton.[9,30-32] It is likely that the exciton binding energy of perovskite depends on kinds of ions forming the perovskite lattice, alloy composition, and impact of screening effect,[7] meaning that we need to carefully identify the types of electronic excited state at room temperature. Furthermore, it has been pointed out that spin-related phenomena, e.g. spin-orbit coupling and/or Rashba splitting effect,[33] cause a conduction band reconstruction and can trigger momentum- and/or spin-forbidden transitions.[34-36] These are relevant to excited-state lifetime and radiative recombination rate, leading to a variation in the lasing threshold. Also, strong reabsorption effect and low nonradiative recombination rate of perovskite may cause a repetitive event of photon emission, reabsorption, and carrier (or exciton) diffusion, known as photon recycling.[37,38] As the photon recycling leads to a spatial diffusion of the excited species in a crystal,[39,40] it may affect the formation of population inversion. However, there is no study on the relationship between the above-mentioned photophysics and the lasing mechanism.

Here we present a series of experimental results on photoluminescence (PL) measurements for all-inorganic perovskite single crystals and their microcavities and show a novel mechanism of coherent emission based on exciton/free-carrier hybrid nature. The microplate of $CsPbBr_3$ single crystal shows a double-PL profile caused by Rashba splitting effect. Pumping fluence dependent PL measurements indicate the emergence of radiative bimolecular recombination of free carriers. In a form of microcavity, on the other hand, angular dependent PL results clearly shows the emergence of polariton lasing, indicating the excitonic nature of $CsPbBr_3$ crystal. The reservoir state, which provides excited species to the polariton state, is attributed to long-lived free carrier. This quite novel mechanism of polariton lasing is expected to have an advantage over the conventional principles in terms of lowering the polariton condensation threshold. These findings shed light on importance of the perovskite materials as not only the lasing medium but also the platform of polaritonic devices.

## Results & Discussion

We obtain $CsPbBr_3$ microplates by antisolvent vapor-assisted crystallization method. Details of the growth procedure are described in the Method section. In short, small amount of perovskite precursor solution (dimethyl sulfoxide, DMSO) is dripped onto a substrate (a silica plate or a distributed Bragg reflector: DBR). We put the sample in poor solvent (acetonitrile, ACN) mist at room temperature [Fig. 1(a)]. One day later, plate-like crystals with lateral size of ~100 μm grow [Fig. 1(b)]. We can control the thickness of microplate by limiting the vertical space with a pair of the plates (i.e. space-limited antisolvent vapor-assisted crystallization method). Surface profile measurements show that the height of plates is in a range of ~0.2 – 1.5 μm [Figs. S1(a) and S1(b) in Supplementary Information]. Clear (100) facet spontaneously appears on the crystal surface, ensuring formation of high quality microcavities. X-ray diffraction analysis shows the [100]-oriented growth of $CsPbBr_3$ microplates with cubic or orthorhombic phase [see section 1 and Fig. S1(c) in Supplementary Information].[40] The absorption spectrum shown in Fig. 1(c) exhibits a band-edge around ~2.3 eV that is comparable to recent studies.[40] A resonant peak found at ~2.4 eV implies an existence of excitonic nature in CsPbBr₃.[17,18]

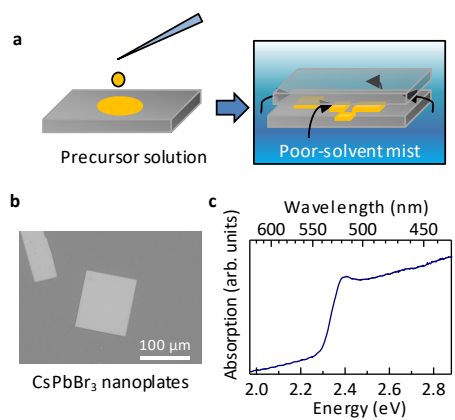

**Figure 1**. CsPbBr₃ single crystal microplates. (a) Schematics for preparing the CsPbBr₃ microplates with anti-solvent vapor-assisted method. Perovskite precursor solution is prepared by solving CsBr and PbBr₂ to DMSO. ACN is used as poor solvent. (b) A scanning electron microscopic image. (c) Absorption spectrum near the band gap energy.

Time-integrated PL spectrum of a bare CsPbBr₃ microplate under optical pulse pumping (~220 nJ/cm² at 405 nm) is shown in Fig. 2(a). The emission spectrum is decomposed into two Gaussian curves centred at ~2.33 and ~2.36 eV (denoted as P1 and P2, respectively), so-called double PL feature. Transient PL dynamics shown in Fig. 2(b) exhibit two decay components having time constants of ~0.67 and 16 ns. The variation of decay profile against the detection energy clearly reveals that the slow and fast decay components are inherent properties of the P1 and P2 emissions, respectively. Some recent studies have claimed that photon-recycling effect results in a red-shifted long-lived emission (i.e. P1 emission).[40,41] As shown in Fig. S2 in Supplementary Information, however, the spectral and temporal PL profiles show no difference between the forward- and back-scattered configurations,[39] revealing a small contribution of the photon-recycling effect in our case, probably due to the crystal thickness as small as ~1 μm. Other studies have shown that strong spin-orbit coupling and/or Rashba splitting effect strongly involve in the near band-edge PL properties and are likely as the origin of double-PL feature.[34-36] It is reported that the strong spin-orbit interaction splits a degenerate exciton state into an optically-passive (dark) singlet state and an optically-active (bright) triplet state.[34,35] These states cannot be observed separately at room temperature as the energies of singlet/triplet separation and triplet splitting (in tetragonal or orthorhombic phases) are within a few meV in the case of $CsPbBr_3$.[34,35] Nevertheless, we can speculate that the PL spectrum of $CsPbBr_3$ microplate involves emissions attributed to optically allowed and forbidden transitions. In addition to this, the Rashba splitting effect also causes the emergence of indirect transition in the momentum space.[36] This would be more prominent for free carrier recombination. Wu et al. have explained very clearly the dynamical Rashba splitting effect as an origin of the double-PL feature,[36] showing that the direct and indirect transitions coexist in the observed PL spectrum. Our results are very similar to their case, i.e. quadratic dependence of initial PL intensity $PL_0$ on pulsed pumping fluence $I_p$. The quadratic dependence emerges in both the long-lived P1 emission and the short-lived P2 emission [Fig. 2(c), below ~5 $\mu J/cm^2$], demonstrating the free-carrier bimolecular recombination. Results of cw excitation measurement shown in the inset of Fig. 2(c) also supports the emergence of free-carrier recombination process; namely cw-PL intensities $PL_{cw}$ show a dependence with the power index of $1.5 - 1.7$ on cw-excitation power density $I_{cw}$, indicating co-existence of radiative bimolecular recombination and nonradiative monomolecular recombination (see subsection 2-3 and Fig. S3 in Supplementary Information). Thus, we conclude that emission from the bare $CsPbBr_3$ microplate involves the direct and indirect transitions (P2 and P1 emissions, respectively) probably due to the Rashba splitting effect, and that the excited species are free carrier.

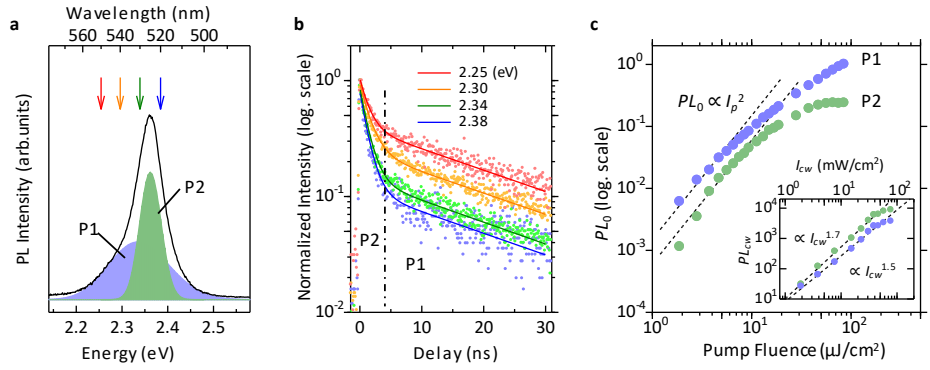

**Figure 2**. PL characterization of a bare CsPbBr₃ microplate with a thickness of ~800 nm. (a) Time-integrated PL spectrum under pumping fluence of ~220 nJ/cm². The spectrum is fitted by two Gaussian curves, P1 and P2, which are shown by blue and green fills, respectively. Arrows indicate detection energies for time-resolved PL measurements. (b) Time-resolved PL decay profiles varied with detection energy. The decay profiles are fitted by bi-exponential functions (solid curves). The slow and fast components correspond to P1 and P2 emission, respectively. (c) Pumping fluence $I_p$ dependence of initial PL intensities $PL_0$ for P1 (blue circles) and P2 (green circles) emissions. Dashed lines show quadratic functions on $I_p$. Inset shows excitation power $I_{cw}$ dependence of steady state PL intensities $PL_{cw}$.

The pumping fluence dependence of $PL_0$ shown in Fig. 2(c) indicates saturation tendencies at $I_p > \sim 10$ μJ/cm². At the even higher $I_p$ ($> \sim 100$ μJ/cm²), the microplate exhibits a lasing action with whispering gallery mode (WGM),[12,13] as shown in Fig. S4 in Supplementary Information. These results imply that most of the excited species are transferred to 'photon-related' excited states, namely the optical cavity modes. This photophysics appears also in 'half-VCSEL' samples that are CsPbBr₃ microplates on a high reflectivity DBR substrate. Figure 3(a) shows a time-integrated PL spectrum for a half-VCSEL sample of a ~860 nm-thick CsPbBr₃ microplate. A new emission peak (P3) appears at the lower energy side of P1 and P2 emissions. The P3 emission shows a thickness-dependent spectral variation as shown in Fig. S5 in Supplementary Information, indicating the optical resonance mode confined in low-Q vertical microcavity as the origin of P3. As the energy range of P3 emission is included in the range of P1 emission, and not significantly overlapped with P2 emission, the P1-releted long-lived excited state provides the excited species to the optical resonance modes (P3 emission). The same photophysics appears more clearly in a 'full-VCSEL' sample. As shown in Fig. 3(b), a below-threshold PL spectrum ($I_p$ ~10 μJ/cm$^2$) of the full-VCSEL sample exhibits only the P3 emission but the P1 and P2 emissions almost disappear. The single-exponential dynamics with a time constant as large as ~94 ns (see inset) demonstrates that the long-lived P1-related state acts as the reservoir of excited species for the P3 emission. As another important point, by comparing Figs. 2(a) and 3(a), we can find that the relative intensity of P2 emission is much smaller in half-VCSEL sample than that in the bare microplate. Furthermore, it decreases monotonically with the increased $I_p$ (see Fig. S6 in Supplementary Information). We deduce from these results that an energy transfer process of P2 (short-lived state)→P1 (long-lived state)→P3 (photon-related state) is facilitated by the high excitation density and the strong optical confinement. We have found in a very recent study that Ruddlesden–Popper layered perovskite microplates and their microcavities exhibit a very similar energy transfer process.[42] The detailed rate equation analyses have revealed that the energy transfer from a short-lived excited state (bright singlet exciton) to a long-lived excited state (dark triplet exciton) is facilitated by the reabsorption of emission.[42] The CsPbBr$_3$ microplate studied here also show similar excitation dynamics, which is established for free carriers involved in the direct and indirect transitions in momentum space.

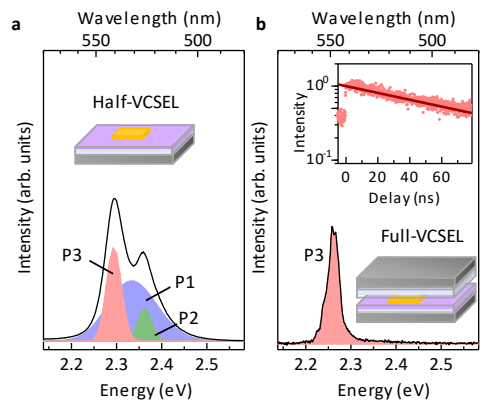

**Figure 3**. Below-threshold PL characterization of VCSEL samples with CsPbBr$_3$

microplates. (a) Time-integrated PL spectrum of a half-VCSEL sample with a crystal thickness of ~860 nm. The pumping fluence is ~700 μJ/cm². The spectrum is fitted by three Gaussian curves, P1 (blue fill, long-lived excited state), P2 (green fill, short-lived excited state), and P3 (red fill, photon-related state) emissions. Inset is a schematic of the half-VCSEL sample. (b) Time-integrated PL spectrum of a half-VCSEL sample with a crystal thickness of ~600 nm. Only the P3 emission is observable. Upper inset shows time-resolved PL profile. Lower inset exhibits a schematic of the full-VCSEL sample.

At the higher $I_p$, the P3 emission of the full-VCSEL sample shows laser-like behaviours, as shown in Fig. S7(a) in Supplementary Information. The energy of laser-like emission depends on the thickness of CsPbBr₃ microplate, but is limited to the energy range of P3 emission due to the energy transfer mechanism discribed above. The clear polarization dependence is also observable [Fig. S7(b)]. Now let us investigate further the underlying physics for this laser-like emission. Figure 4(a) compares emission spectra of a full-VCSEL sample with a crystal thickness of ~1.1 μm measured under three different pumping conditions. At the weak excitation by a cw laser diode, the sample shows a broad emission spectrum centred at ~2.34 eV (see dashed black curve). This energy position corresponds to the P2 emission rather than the P1 emission, showing that the excitation density is not high enough and that the emission from the short-lived excited state (P2-related state) is still dominant. Under the pulsed pumping with $I_p$ of ~180 μJ/cm² (see purple curve), a spectrally narrowed emission appears at ~2.28 eV, which is within the energy range of P3 emission. When $I_p$ increases to ~1.2 mJ/cm², the sample exhibits a laser-like emission (see red curve). The emission intensity shows a S-shaped nonlinear dependence on the pumping fluence with a threshold of ~250 μJ/cm², as shown by red circles in Fig. 4(b). At the same time, the spectral width drastically decreases from ~32 meV to a few meV (see blue triangles). Figures 4(c) – 4(e) are contour maps of angular-resolved PL results ($0 \leq \theta \leq 60$ º) at the three different pumping conditions. Under the cw excitation [Fig. 4(c)], four or five dispersion modes appear around the energy range of P2 emission, which is close to the exciton absorption peak ($E_{ex}$ ~2.4 eV, dashed line). Angular dispersions of these emission modes are well described as the uncoupled cavity-photon modes represented as

$E_{ph}(\theta) = E_{ph}(0)\left(1 - \sin^2\theta/n_{eff}^2\right)^{-1/2}$,[23] where we consider the wavelength dispersion of

background refractive index to account for the effective refractive index ($n_{eff}$ ~1.5 – 2.0, see

subsection 3-1 in Supplementary Information). Under the pulsed pumping, as shown in Fig. 4(d), the

emission range shifts to the range of P1 emission (~2.25 – 2.34 eV). More importantly, the

uncoupled cavity-photon modes tend to disappear, and another mode emerges instead. The angular

dispersion of the newly observed mode, $E_{pol}(\theta)$, is well explained by a simple coupled oscillator

model formed with $E_{ex}$ and $E_{ph}(\theta)$,[22,23] as shown by a thick dashed curve in Fig. 4(d) (see

subsection 3-2 and Fig. S8 in Supplementary Information). This result clearly reveals a formation of

cavity polariton mode. The Rabi splitting energy of ~270 meV is enough larger than the linewidths

of $E_{ex}$ and $E_{ph}$ (those are estimated to be ~50 meV for both in the absorption peak and the below-

threshold emission peak, respectively). Thus the microcavity system is under strong coupling

regime.[22,23] At the higher pumping fluence, as shown in Fig. 4(e), the emission signal eventually

condenses into the energy minimum of the polariton dispersion curve, demonstrating the polariton

lasing at room temperature.[17,18,43-48] The above-threshold emission exhibits a blue shift in its peak

energy and a slight increase in linewidth [see green squares and blue triangles in Fig. 4(b)]. These

features are due to the polariton-polariton interaction in the condensed polariton state.[24] We have

confirmed the reproducibility of these observations as shown in Figs. S9 and S10 in Supplementary

Information. While there is a variation in the threshold fluence and the dispersion characteristics, all samples with different crystal thicknesses show the same features as in Fig. 4, that are the transition in dispersion characteristic and the condensation into energy minima.

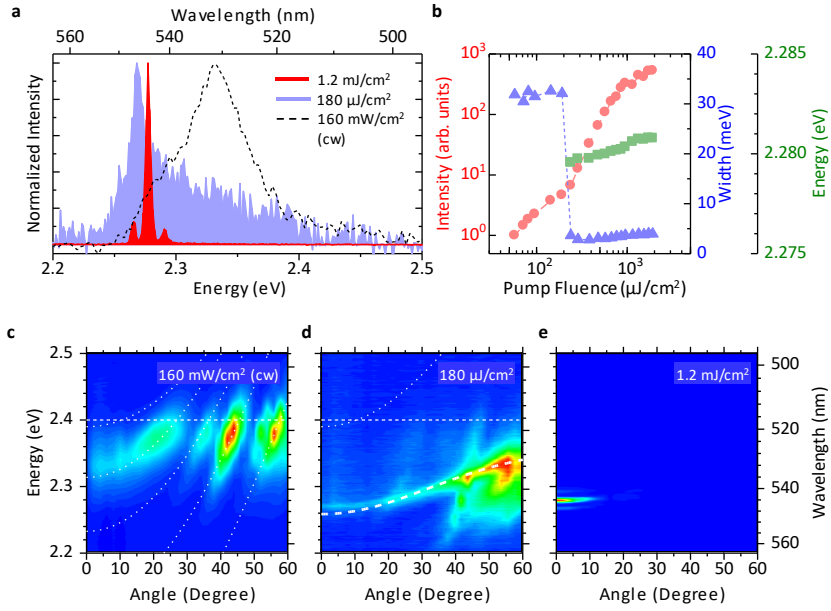

**Figure 4**. Polariton lasing in a full-VCSEL of CsPbBr₃ microplate with a thickness of ~1.1 μm. (a) Comparison of PL spectra ($\theta$ = 0 °) obtained under cw-160 mW/cm² excitation, 180 μJ/cm², and 1.2 mJ/cm² pulsed pumping. (b) Pumping fluence dependencies of intensity (pink circles), line width (blue triangles), and peak position (green squares). (c) – (e) Contour maps of angle-dependent PL spectra at different pumping conditions. (c) The map obtained at the cw-160 mW/cm² excitation. Dotted curves and a dashed line exhibit uncoupled cavity photon modes and exciton mode, respectively. (d) The map obtained at the pulsed-180 μJ/cm² pumping (below threshold). A thick dashed curve shows a lower polariton branch resulting from strong coupling between cavity photon mode (dotted curve) and exciton mode (dashed line). (e) The map obtained at the pulsed-1.2 mJ/cm² pumping (above threshold). Energy condensation attributed to polariton lasing is clearly observed.

The photon-exciton strong coupling and polariton lasing reveal a presence of excitonic nature in CsPbBr₃. This fact contradicts the observation of free carrier-like PL characteristics described before. Figure 5 depicts a physical model explaining the polariton lasing behaviour based on the exciton/free-carrier hybrid nature. At the weak pumping fluence [see Fig. 5(a)], hot carriers generated by the nonresonant optical pumping first cool down to the P2-related state with a large radiative recombination rate. The radiative coupling from the P2-related state to the uncoupled cavity photon modes emerges preferentially, as observed in Fig. 4(c). This is quite reasonable given the Purcell effect.[49] The system is thus in the weak coupling regime. As the pumping fluence increases, the physical picture changes dramatically, as shown in Fig. 5(b). The energy transfer from the short-lived P2-related state to the long-lived P1-related state is facilitated. Then the free carriers in the P1-related state can radiatively couple to the polariton branch, as observed in Fig. 4(d). The transition between the weak and strong coupling regimes would come from the difference in the lifetimes of P1- and P2-related states; namely, the coupling to the polariton mode requires the high excitation density in the reservoir state. It should be noted that, as found in Fig. 4(d), the polariton population below the threshold is prominent around the bottleneck position ($\theta$ = 50 – 60 º). This fact implies that the stimulated cooling mechanism for the polariton condensation is due to the polariton-polariton scattering, indicating again the excitonic nature of CsPbBr$_3$.

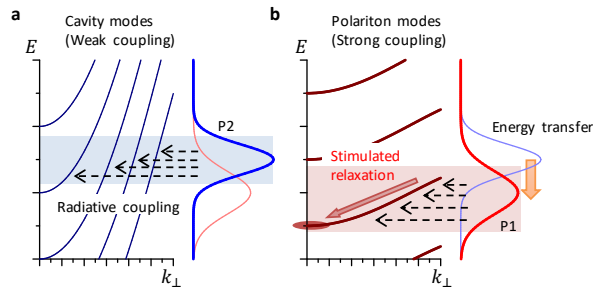

**Figure 5**. Schematic diagram illustrating relaxation and emission models in CsPbBr$_3$ microcavity. (a) A model for the low-pumping regime (below threshold). The microcavity system is in weak coupling regime. The excited species (free carriers) in the short-lived P2-related mode radiatively couple to bare cavity photon modes. (b) A model for the high pumping fluence regime (above threshold). The excited species are transferred to the long-lived P1-related mode. The system exhibits a strong coupling regime where the radiative coupling from the P1-related state to the polariton mode emerges.

In the discussion above, a question remains whether the main excited species is an exciton or a free carrier. The exciton binding energies of bromide-based perovskites have been reported to be 30 – 60 meV.[32] This is close to the activation energy at room temperature and are relatively large compared to those of iodide-based perovskites. Stranks et al. have proposed a model describing equilibrium properties between free carriers and excitons.[31] Given those studies, the excitons and free carriers in $CsPbBr_3$ may coexist and/or hybridize at room temperature. Nevertheless, it is interesting to note that the polariton lasing has been reported only in all-inorganic perovskite, i.e. $CsPbBr_3$ or $CsPbCl_3$,[17,18] whereas hybrid organic-inorganic halide perovskites do not show the strong coupling or polariton lasing.[49,50] We also examine a similar experiment for a full-VCSEL sample with $MAPbBr_3$ (MA: methylammonium) microplate and find no polariton-like behaviour in the lasing action (see Fig. S11 in Supplementary Information). This fact clearly reveals the absence of excitonic nature in $MAPbBr_3$.

## Conclusions

Our experimental results demonstrate that the hybrid nature of $CsPbBr_3$ leads to the unprecedented lasing mechanism, i.e. a polariton lasing based on free carrier reservoir. This is clearly distinguishable from the conventional polariton lasing where the (dark) exciton states serve as the excitation reservoir. Of course, it is quite different from the conventional photon lasing. We can expect some crucial advantages in the novel mechanism; (i) the reservoir state with a long lifetime and a large quantum yield is advantageous for lowering the threshold of polariton condensation, (ii) deactivation of polaritons by the Auger process is avoidable owing to the difference in momentum dispersion between the electron and polariton, and (iii) the free carrier nature in reservoir state is promising for the current driving. These features pave the way to develop a platform of room temperature polariton devices that are used not only for the highly functional coherent light sources but also for the low-power consumption quantum information technologies.

## Acknowledgements

K.Y. acknowledge funding from Japan Society for the Promotion of Science, JSPS KAKENHI (No. 18H01476). The authors thank to Prof. M. Nakayama and T. Obuchi for their help on microcavity preparations.

## Author contributions

S.E. and K.Y. conceived and planned the experiments. S.E. fabricated measurement samples. S.E. and K.F. performed PL measurements. T.T. set up angular dependent measurement system. Y.U. performed designing microcavity devices. S.E., S.T.., and K.Y. drafted the manuscript and complied figures, with discussion of results and feedback from all authors.

## Competing interest

The authors declare no competing interests.

## References

1.  Stranks, S. D., and Snaith, H. J. Metal-halide perovskites for photovoltaic and light-emitting devices. *Nat. Nanotechnol.* **10**, 391-402 (2015).

2.  Green, M. A., Ho-Baillie, A., and Snaith, H. J. The emergence of perovskite solar cells. *Nat. Photon.* **8**, 506-514 (2014).

3.  Sutherland, B. R., and Sargent, E. H. Perovskite photonic sources. *Nat. Photon.* **10**, 295-302 (2016).

4.  Tan, Z.-K., et al. Bright light-emitting diodes based on organometal halide perovskite. *Nat. Nanotechnol.* **9**, 687-692 (2014).

5.  Liu, Y., et al. Efficient blue light-emitting diodes based on quantum-confined bromide perovskite nanostructures. *Nat. Photon.* **13**, 760-764 (2019).

6.  Deschler, F., et al. High photoluminescence Efficiency and optically pumped lasing in solution-processed mixed halide perovskite semiconductors. *J. Phys. Chem. Lett.* **5**, 1421-1426 (2014).

7.  Quan, L. N., F. García de Arquer, P., Sabatini, R. P., and Sargent, E. H. Perovskites for light emission. *Adv. Mater.* **30**, 1801996 (2018).

8.  Xing, G., et al. Low-temperature solution-processed wavelength-tunable perovskites for lasing. *Nat. Mater.* **13**, 476-480 (2014).

9.  Chen, S., and Nurmikko, A. Excitonic gain and laser emission from mixed-cation halide perovskite thin films. *Optica* **5**, 1141-1149 (2018).

10. Brenner, P., et al. Highly stable solution processed metal-halide perovskite lasers on nanoimprinted distributed feedback structures. *Appl. Phys. Lett.* **109** 141106 (2016).

11. Whitworth, G. L., et al. Nanoimprinted distributed feedback lasers of solution processed hybrid perovskites. *Opt. Express* **24**, 23677-23684 (2016).

12. Zhang, Q., Ha, S. T., Liu, X., Sum, T. C., and Xiong, Q. Room-temperature near-infrared high-Q perovskite whispering-gallery planar nanolasers. *Nano Lett.* **14**, 5995-6001 (2014).

13. Zhu, H., et al. Lead halide perovskite nanowire lasers with low lasing thresholds and high quality factors. *Nat. Mater.* **14**, 636-642. (2015).

14. Kurahashi, N., Nguyen, V.-C., Sasaki, F., and Yanagi, H. Whispering gallery mode lasing in lead halide perovskite crystals grown in microcapillary. *Appl. Phys. Lett.* **113**, 011107 (2018).

15. Wang, Y., et al. All-inorganic colloidal perovskite quantum dots: A new class of lasing materials with favorable characteristics. *Adv. Mater.* **27**, 7101-7108 (2015).

16. Akkerman, Q. A., Rainò, G., Kovalenko, M. V., and Manna, L. Genesis, challenges and opportunities for colloidal lead halide perovskite nanocrystals. *Nat. Mater.* **17**, 394-405 (2018).

17. Su, R., et al. Room-temperature polariton lasing in all-inorganic perovskite nanoplatelets. *Nano Lett.* **17**, 3982-3988 (2017).

18. Bao, W., et al. Observation of Rydberg exciton polaritons and their condensate in a perovskite cavity. *Proc. Natl. Acad. Sci. U.S.A.* **116**, 20274-20279 (2019).

19. Weisbuch, C., Nishioka, M., Ishikawa, A., and Arakawa, Y. Observation of the coupled exciton-photon mode splitting in a semiconductor quantum microcavity. *Phys. Rev. Lett.* **69**, 3314-3317 (1992).

20. Lidzey, D. G., et al. Strong exciton-photon coupling in an organic semiconductor microcavity. *Nature* **395**, 53-55 (1998).

21. Goto, K., Yamashita, K., Yanagi, H., Yamao, T., and Hotta, S. Strong exciton-photon coupling in organic single crystal microcavity with high molecular orientation. *Appl. Phys. Lett.* **109**, 061101 (2016).

22. Deng, H., Haug, H., and Yamamoto, Y. Exciton-polariton Bose-Einstein condensation. *Rev. Mod. Phys.* **82**, 1489-1537 (2010).

23. Kavokin, A. V., Baumberg, J. J., Malpuech, G. & Laussy, F. P. Microcavities (Oxford University Press, 2017).

24. Kasprzak, J., et al. Bose–Einstein condensation of exciton polaritons. *Nature* **443**, 409-414 (2006).

25. Sanvitto, D., and Kéna-Cohen, S. The road towards polaritonic devices. *Nat. Mater.* **15**, 1061-1073 (2016).

26. Berlo, N. G., et al. Realizing the classical XY Hamiltonian in polariton simulators. *Nat. Mater.* **16**, 1120-1126 (2016).

27. Zasedatelev, A. V., et al. A room-temperature organic polariton transistor. *Nat. Photon.* **13**, 378-383 (2019).

28. Birnbaum, K. M., et al. Photon blockade in an optical cavity with one trapped atom. *Nature* **436**, 87-90 (2005).

29. Wei, Q., et al. Recent progress in metal halide perovskite micro- and nanolasers. *Adv. Opt. Mater.* **7**, 1900080 (2019).

30. D'Innocenzo, V., et al. Excitons versus free charges in organo-lead tri-halide perovskites. *Nat. Comuun.* **5**, 3586 (2014).

31. Stranks, S. D., et al. Recombination kinetics in organic-inorganic perovskites: excitons, free charge, and subgap states. *Phys. Rev. Appl.* **2**, 034007 (2014).

32. Jiang, Y., Wang, X., and Pan, A. Properties of excitons and photogenerated charge carriers in metal halide perovskites. *Adv. Mater.* **22**, 1806671 (2019).

33. Kepenekian, M., et al. Rashba and Dresselhaus effects in hybrid organic-inorganic perovskites: from basics to devices. *ACS Nano* **9**, 11557-11567 (2015).

34. Becker, M. A., et al. Bright triplet excitons in caesium lead halide perovskites. *Nature* **553**, 189-193 (2018).

35. Tamarat, P., et al. The ground exciton state of formamidinium lead bromide perovskite nanocrystals is a singlet dark state. *Nat. Mater.* **18**, 717-724 (2019).

36. Wu, B., et al. Indirect tail states formation by thermal-induced polar fluctuations in halide perovskites. *Nat. Commun.* **10**, 484 (2019).

37. Yamada, Y., Nakamura, T., Endo, M., Wakamiya, A., and Kanemitsu, Y. Photocarrier recombination dynamics in perovskite $CH_3NH_3PbI_3$ for solar cell applications. *J. Am. Chem. Soc.* **136**, 11610-11613 (2014).

38. Pazos-Outón, L. M., et al. Photon recycling in lead iodide perovskite solar cells. *Science* **351**, 1430-1433 (2016).

39. Fang, Y., Wei, H., Dong, Q., and Huang, J. Quantification of re-absorption and re-emission processes to determine photon recycling efficiency in perovskite single crystals. *Nat. Commun.* **8**, 14417 (2017).

40. Dursun, I., et al. Efficient photon recycling and radiation trapping in cesium lead halide perovskite waveguides. *ACS Energy Lett.* **3**, 1492-1498 (2018).

41. Yamada, Y., Hoyano, M., Akashi, R., Oto, K., and Kanemitsu, Y. Impact of chemical doping on optical responses in bismuth-doped $CH_3NH_3PbBr_3$ single crystals: carrier lifetime and photon recycling. *J. Phys. Chem. Lett.* **8**, 5798-5803 (2017).

42. Fujiwara, K., et al. Excitation dynamics in layered lead halide perovskite crystal slabs and microcavities. *ACS Photon.* 10.1021/acsphotonics.0c00038 (2020).

43. Christopoulos, S., et al. Room-temperature polariton lasing in semiconductor microcavities. *Phys. Rev. Lett.* **98**, 126405 (2007).

44. Kéna-Cohen, S., and Forrest, S. R. Room-temperature polariton lasing in an organic single-crystal microcavity. *Nat. Photon.* **4**, 371-375 (2010).]

45. Plumhof, J. D., Stöferle, T., Mai, L., Scherf, U., and Mahrt, R. F. Room-temperature Bose–Einstein condensation of cavity exciton–polaritons in a polymer. *Nat. Mater.* **13**, 247-252 (2014).

46. Daskalakis, K. S., Maier, S. A., Murray, R., and Kéna-Cohen, S. Nonlinear interactions in an organic polariton condensate. *Nat. Mater.* **13**, 271-278 (2014).

47. Tanaka, Y., et al. Vertical cavity lasing from melt-grown crystals of cyano-substituted thiophene/phenylene co-oligomer. *Appl. Phys. Lett.* **107**, 163303 (2015).

48. Yamashita, K., et al. Ultrafast dynamics of polariton cooling and renormalization in an organic single-crystal microcavity under nonresonant pumping. *ACS Photon.* **5**, 2182-2188 (2018).

49. Wang, J., et al. Purcell effect in an organic-inorganic halide perovskite semiconductor microcavity system. *Appl. Phys. Lett.* **108**, 022103 (2016).

50. Bouteyre, P., et al. Room-temperature cavity polaritons with 3D hybrid perovskite: toward large-surface polaritonic devices. *ACS Photon.* **6**, 1804-1811 (2019).

## Methods

*Preparation of Precursor Solutions*

CsBr and PbBr$_2$ were purchased from Tokyo Chemical Industry and used as received. They were dissolved in DMSO at a concentration of 0.4 M. The solution was stirred at 300 rpm and 50 ºC for 10 minutes. After that, ACN was added as a poor solvent until the solution reached saturation. The solution was stirred at room temperature for 3 hours and subsequently filtered.

*Crystal Growth*

CsPbBr$_3$ microplates were grown by antisolvent vapor-assisted crystallization method.[51,52] Two silica plates (area: $10 \times 10$ mm$^2$ and thickness 0.5 mm) were cleaned with acetone, ethanol, and isopropyl alcohol in an ultrasonic bath for 5 minutes for each solvent. One of the plates was exposed to UV ozone to obtain a hydrophilic surface. The other was spin coated with hexamethyldisilazane in a two-step sequential condition (500 rpm for 5 seconds followed by 3,000 rpm for 10 seconds) to obtain a hydrophobic surface. 10 µl of the perovskite precursor solution mentioned above was casted on the hydrophilic substrate. The hydrophobic substrate was put on the solution, and then the two plates were pinched with a bulldog clip. The sample was placed in a glass beaker (100 ml), into which 3 ml of ACN was dropped. The glass beaker was sealed with cling film and was put on a hotplate (40 ºC) to fill the inner space of beaker with the ACN mist. The ACN mist penetrated the gap between the two plates slowly, promoted the nucleation of CsPbBr$_3$, resulting in microplate after one day. We could control the thickness of microplate by limiting the vertical space with a pair of the plates (i.e. space-limited antisolvent vapor-assisted crystallization method).[51-54] The lateral size and thicknesses of crystals were evaluated by a scanning electron microscope (TM3030 plus, Hitachi) and a surface profiler (Dektak XT-S, Bruker) observations, respectively, after removing the top plate. Typically, the microplates showed square or rectangular shapes with a dimension of 50 – 200 µm, and their thickness is in a range of 200 nm – 3.5 µm. X-ray diffraction measurements (D8 Discover, Bruker) showed two distinct peaks at 15.2 º and 30.5 º, which are attributed to diffractions of (100) and (200) planes, respectively, and consistent with a recent report on the cubic or orthorhombic CsPbBr$_3$ microplate.[40,51,55]

*Microcavity Fabrication*

To fabricate microcavities, glass plates on which DBRs were deposited were used instead of the untreated silica plates. The DBRs were multilayers of SiO$_2$ and TiO$_2$ on BK7 glass plates (area: $10 \times 10$ mm$^2$ and thickness 0.5 mm). The rf-magnetron sputtering method was employed for the deposition of nine pairs of SiO$_2$ and TiO$_2$ layers. The resultant DBR had a reflectivity larger than 99 %. In this study we used two types of DBRs [see Fig. 4(a)]. One of them had a reflection band of ~450 – 550 nm and was used as the top mirror. The other had a reflection band of ~500 – 600 nm and was used as the bottom mirror. This combination enabled the formation of microcavity with a band of 2.1 – 2.6 eV as well as good transparency for excitation light (405 nm or 351 nm). See Fig. S7 in the Supplementary Information for more detailed.

*Optical Characterizations*

All optical measurements were carried out at room temperature. The absorption measurement was performed by using a spectrometer (UV-2500PC, Shimadzu). For weakly pumped PL measurements, we used a 405-nm diode laser with a pulse with of < 39 ps as excitation light source (PiL040X, A. L. S. GmbH), which operated at repetition frequency of 10 MHz and output power of 200 µW. Time-integrated PL spectra were recorded by using a monochromator (Triax 550, Horiba) equipped with a multichannel photodetector (Synapse, Horiba). The spectral resolution was ~0.3 nm. The emission counts were measured with time-integration for 1 s. To perform time-resolved PL measurements, the emission was dispersed in a monochromator (SPG-120SS, SHIMADU) and detected by a single-photon detector (ID-100, ID Quantique). The signal was collected with a time-correlated single photon counting module (SPC-130EM, Becker & Hickl GmbH). Neutral density filters were used for varying the pumping fluence. For strongly pumped PL measurements, the third harmonic generation of the Nd: YLF ~8-ns pulse laser with a wavelength of 351 nm and a repetition rate of < 10 Hz (Quantas-Q1D-TH-ATF, Quantum Light Instruments) was used as an excitation source. In the angle-resolved PL measurements, the laser light was focused using a plano-convex lens with a focal length of 200 mm. The detection angle of emission was varied by using a home-made rotational stage on which an optical fibre (core diameter of 1 mm) for collecting the emission was placed.

References

51. Yang, Z., et al. Large and Ultrastable all-inorganic CsPbBr$_3$ monocrystalline films: low-temperature growth and application for high-performance photodetectors. *Adv. Mater.* **30**, 1802110 (2018).

52. Wang, X.-D., Li, W.-G., Liao, J.-F., and Kuang, D.-B. Recent advances in halide perovskite single-crystal thin films: fabrication methods and optoelectronic applications. *Sol. PRL* **3**, 1800294 (2019).

53. Yu, W., et al. Single crystal hybrid perovskite field-effect transistors. *Nat. Commun.* **9**, 5354 (2018).

54. Han, L., Liu, C., Wu, L., and Zhang, J. Observation of the growth of MAPbBr$_3$ single-crystalline thin film based on space-limited method. *J. Cryst. Growth* **501**, 27-33 (2018).

55. Lan, S., et al. Vapor-phase growth of CsPbBr$_3$ microstructures for highly efficient pure green light emission. *Adv. Opt. Mater.* **7**, 1801336 (2019).